\documentclass[aps,reprint,twocolumn,showpacs]{revtex4}
\usepackage{amsmath}
\usepackage{epsfig}
\usepackage{dcolumn}
\usepackage{bm}
\usepackage{amsfonts}
\usepackage{graphicx}
\usepackage{color}

\begin{document}

\title{Charge transfer statistics and qubit dynamics at tunneling Fermi-edge
singularity.}
\author{V.V. Ponomarenko$^{1}$ and I. A. Larkin$^{2}$}
\affiliation{$^1$Nonlinearity and Complexity Research Group, Aston University, Birmingham
B4 7ET, United Kingdom \\
$^2$Institute of Microelectronics Technology RAS, 142432 Chernogolovka,
Russia}
\date{\today }

\begin{abstract}
Tunneling of spinless electrons from a single-channel emitter into an empty
collector through an interacting resonant level of the quantum dot (QD) is
studied, when all Coulomb screening of charge variations on the dot is
realized by the emitter channel and the system is mapped onto an exactly
solvable model of a dissipative qubit. In this model we describe the qubit
density matrix evolution with a generalized Lindblad equation, which permit
us to count the tunneling electrons and therefore relate the qubit dynamics
to the charge transfer statistics. In particular, the coefficients of its
generating function equal to the time dependent probabilities to have the
fixed number of electrons tunneled into the collector are expressed through
the parameters of a non-Hermitian Hamiltonian evolution of the qubit pure
states in between the successive electron tunnelings. From the long time
asymptotics of the generating function we calculate Fano factors of the
second and third order (skewness) and establish their relation to the extra
average and cumulant, respectively, of the charge accumulated in the
transient process of the empty QD evolution beyond their linear time
dependence. It explains the origin of the sub and super Poisson shot noise
in this system and shows that the super Poisson signals existence of a
non-monotonous oscillating transient current and the qubit coherent
dynamics. The mechanism is illustrated with particular examples of the
generating functions, one of which coincides in the large time limit with
the $1/3$ fractional Poissonian realized without the real fractional charge
tunneling.
\end{abstract}

\maketitle

\section{Introduction}

The Fermi-edge singularity (FES) resulting \cite{1,2} from the
reconstruction of the Fermi sea of conduction electrons under a sudden
change of a local potential have been primarily observed \cite{3,4} as a
power-law singularity in X-ray absorption spectra. A similar phenomenon of
the FES in transport of spinless electrons through a quantum dot (QD) was
predicted \cite{5} in the perturbative regime when a localized QD level is
below the Fermi level of the emitter in its proximity and the collector is
effectively empty (or in equivalent formulation through the particle-hole
symmetry) and the tunneling rate of the emitter is sufficiently small. Then,
the subsequent separated in time electron tunnelings from the emitter vary
the localized level charge and generate sudden changes of the scattering
potential leading to the FES in the I-V curves at the voltage threshold
corresponding to the resonance. Direct observation of these perturbative
results in experiments \cite{geim,6,7,8,9,lar}, however, is complicated due
to the uncontrolled effects such as of a finite life time of the electrons
(the level broadening of the QD localized state), temperature smearing and
variation of tunneling parameters due to application of the bias voltage.
Therefore, it has been suggested \cite{epl} that the true FES nature of a
threshold peak in the I-V dependence can be verified through observation of
the oscillatory behavior of a corresponding time-dependent transient
current. Indeed, in the FES theory \cite{1,2} appearance of such a threshold
peak signals formation of a two-level system of the exciton electron-hole
pair or qubit in the tunneling channel at the QD. The qubit undergoes
dissipative dynamics characterized \cite{L,Sch}, in the absence of the
collector tunneling, by the oscillations of the levels occupation. It should
create an oscillating transient current at least for a weak enough collector
tunneling rate. Although a direct observation of these oscillations would
give the most clear verification of the nature of the I-V threshold peaks,
it involves measurement of the time dependent transient current averaged
over its quantum fluctuations, which is a challenging experimental task. In
the recent experiments \cite{8,noise,noise2} the low-temperature short noise
measurements have been carried out for this purpose. These measurements
showed existence of the sub and super Poisson statistics of the current
fluctuations at the FES and have raised \cite{8,noise} a new interest \cite%
{novotny} in the qubit dynamics, though their coherency manifestation in the
current fluctuations needs to be further clarified. Also, for this purpose
the methods of measurement of the third order current cumulants \cite%
{shovkun, third} could be considered.

Therefore, in this work we study quantum fluctuations of the charge
transferred into collector and their reflection of the coherent qubit
dynamics in the FES regime in the simplified, but still realistic setup
suggested earlier \cite{epl,prb}, in which all Coulomb screening of sudden
charge variations on the QD during the spinless electron tunneling is due to
a single tunneling channel of the emitter. It can be realized, in
particular, if the emitter is represented by a single edge-state in the
integer quantum Hall effect. This system is described by a non-equilibrium
model of an interacting resonant level, which can be mapped \cite{epl} onto
an exactly solvable model of a dissipative qubit. Making use of its solution
it was demonstrated earlier that FES in the I-V dependence in this system is
accompanied for a wide range of the model parameters by an oscillating
behavior \cite{epl} of the collector transient current, in particular, when
the QD evolves from its empty state and that the qubit dynamics also
manifest themselves through the resonant features of the \textit{a.c.}
response \cite{prb}.

Here we further study quantum fluctuations of the charge transfer in this
model by applying the method of full counting statistics \cite{lll,naz}. For
this purpose we derive a generalized Lindblad equation, which describes the
qubit density matrix evolution and simultaneouly permit us to count the
tunneling electrons and therefore relate the qubit dynamics to the charge
transfer statistics. From this equation it follows that the generating
function of the charge transfer in an arbitrary evolution process can be
expressed through the generating function for the process initiating from
the empty QD. The latter is found by dividing the whole process of the qubit
evolution into separate time intervals between the successive electron
tunnelings. In these intervals dynamics of the qubit pure states are
governed by a non-Hermitian Hamiltonian. The coefficients of the generating
function equal to the time dependent probabilities to have the corresponding
fixed number of electrons tunneled into the collector are determined by
matrix elements of the non-Hermitian Hamiltonian evolution operator and
therefore can be used to extract from them the parameters of this evolution
(the frequencies and the damping rates).

From the linear in time part of the long time asymptotics of the generating
function logarithm or cumulant generating function (CGF) we calculate the
zero-frequency reduced current correlators commonly studied in the full
counting statistics. Normalized by the average stationary current, the
second order correlator known as Fano factor $F_2$ predicts existence in
this system the parametrical regions of the sub Poisson short noise $F_2<1$
around the resonance and the super Poisson noise $F_2>1$ far from the
resonance. Among the generalized higher order Fano factors given by the
normalized higher order current correlators we examine the third one called
skewness and find a small parametric area, where it changes its sign and
becomes negative.

We also study the next order finite term of the CGF long time asymptotics,
which determines the extra charge accumulation in the transient process, in
particular, of the empty QD evolution beyond the one characterized by the
linear in time cumulants. We establish a direct relation between the Fano
factor and the extra charge average and between the skewness and the extra
charge cumulants. This relation explains the origin of the sub and super
Poisson shot noise in this system and shows that the super Poissonian means
existence of a non-monotonous oscillating transient current as a consequence
of the qubit coherent dynamics.

The mechanism is illustrated with particular examples of the generating
functions in the special regimes, one of which coincides in the large time
limit with the $1/3$ fractional Poissonian realized without the real
fractional charge tunneling. This example underlines that observation of the
fractional charge in the Poissonian short noise is necessary, but not
sufficient to prove its real tunneling.

The paper is organized as follows. In Sec. II we introduce the model and
formulate those conditions, which make it solvable through a standard
mapping onto the dissipative two-level system or qubit. In Sec. III we apply
the non-equilibrium Keldysh technique to derive the generalized Lindbladian
equation describing the dissipative evolution of the qubit density matrix
and counting the charge transferred into the collector. Its properties are
studied. In particular, we find its stationary solution and the stationary
tunneling current and derive a simple relation between the generating
functions of the charge transfer during the two processes initiating from
the empty and stationary state of the QD as a special case of the general
expression for generating function for an arbitrary evolution process
through the one for the process initiating from the empty QD.

In Sec. IV we consider the non-Hermitian Hamiltonian evolution of the qubit
two-level system in between the successive electron tunneling. Both
two-level energies modified by the collector tunneling rate acquire in
general different imaginary parts. We find the evolution operator and use
its matrix elements to calculate the generating function for the empty QD
evolution. Its coefficients are studied to relate the time dependence of
probabilities to find the corresponding fixed number of electrons tunneled
into the collector to the qubit evolution.

In Sec. V we calculate the zero-frequency reduced current correlators (or
current cumulants) defined by the leading exponent of the generating
function independent of the QD initial state and discuss behavior of the
Fano factor and skewness. We also find the extra average and second order
cumulant of the charge accumulated in the process of the empty QD evolution
which are defined by the prefactor of the leading exponent. It turns out
that the Fano factors and the extra charge moments are not independent. To
establish connection between them we make use of the above relation between
the two generating functions.

In Section VI two generating functions are calculated asymptotically in the
two regimes when amplitude of the qubit two-level coupling is much smaller
than the collector tunneling rate or the absolute value of the QD level
energy and in the opposite limit when the amplitude is much larger than both
of them. Accumulation of the extra charge in these regimes is illustrated
with the corresponding transient current behavior. We also calculate the
generating function at the special point of degeneracy of the two qubit
levels energies including their imaginary parts. We find that in this
special case it takes the $1/3$ fractional Poissonian form, where all
probabilities of tunneling of the fractional charges mean tunneling of the
charges integer parts. The large time limit of this function nontheless
coincides with the true $1/3$ fractional Poisson. This example underlines
that observation of the fractional charge in the Poissonian short noise is
necessary, but not sufficient to prove its real tunneling. The results of
the work are summarized in the Conclusion.

\section{Model}

The system we consider below is described with Hamiltonian $\mathcal{H}=%
\mathcal{H}_{res}+\mathcal{H}_{C}$ consisting of the one-particle
Hamiltonian of resonant tunneling of spinless electrons and the Coulomb
interaction between instant charge variations of the dot and electrons in
the emitter. The resonant tunneling Hamiltonian takes the following form
\begin{equation}
\mathcal{H}_{res}=-\epsilon _{d}d^{+}d+\sum_{a=e,c}\mathcal{H}_{0}[\psi
_{a}]+w_{a}(d^{+}\psi _{a}(0)+h.c.)\ ,  \label{hres}
\end{equation}%
where the first term represents the resonant level of the dot, whose energy
is $-\epsilon _{d}$. Electrons in the emitter (collector) are described with
the chiral Fermi fields $\psi _{a}(x),a=e(c)$, whose dynamics is governed by
the Hamiltonian $\mathcal{H}_{0}[\psi ]==-i\!\!\int \!dx\psi ^{+}(x)\partial
_{x}\psi (x)\ (\hbar =1)$ with the Fermi level equal to zero or drawn to $%
-\infty $, respectively, and $w_{a}$ are the correspondent tunneling
amplitudes. The Coulomb interaction in the Hamiltonian $\mathcal{H}$ is
introduced as
\begin{equation}
\mathcal{H}_{C}=U_{C}\psi _{e}^{+}(0)\psi _{e}(0)(d^{+}d-1/2)\ .  \label{hc}
\end{equation}%
Its strength parameter $U_{C}$ defines the scattering phase variation $%
\theta $ for electrons in the emitter channel and therefore the change of
the localized charge in the emitter $\delta n=\theta /\pi \ \ (e=1)$, which
we assume provides the perfect screening of the QD charge: $\delta n=-1$.

After implementation of bosonization of the emitter Fermi field $\psi
_{e}(x)=\sqrt{\frac{D}{2\pi }}\eta e^{i\phi (x)}$, where $\eta $ denotes an
auxiliary Majorana fermion, $D$ is the large Fermi energy of the emitter,
and the chiral Bose field $\phi (x)$ satisfies $[\partial _{x}\phi (x),\phi
(y)]=i2\pi \delta (x-y)$, and further completion of a standard rotation \cite%
{schotte}, under the above screening assumption we have transformed \cite%
{epl} $\mathcal{H}$ into the Hamiltonian of the dissipative two-level system
or qubit:
\begin{eqnarray}
\mathcal{H}_{Q} &=&-\epsilon _{d}d^{+}d+\mathcal{H}_{0}[\psi
_{c}]+w_{c}(\psi _{c}^{+}(0)e^{i\phi (0)}d+h.c.)  \notag \\
&&+\Delta \eta (d-d^{+})\ ,  \label{hq}
\end{eqnarray}%
where $\Delta =\sqrt{\frac{D}{2\pi }}w_{e}$ and the time dependent
correlator of electrons in the empty collector $\langle \psi _{c}(t)\psi
_{c}^{+}(0)\rangle =\delta (t)$ will allow us to drop the bosonic exponents
in the third term on the right-hand side in (\ref{hq}).

\section{Lindblad equation for the qubit evolution and count of tunneling
charge}

We use this Hamiltonian to describe the dissipative evolution of the qubit
density matrix $\rho _{a,b}(t)$, where $a,b=0,1$ denote the empty and filled
levels, respectively. In the absence of the tunneling into the collector at $%
w_{c}=0$, $\mathcal{H}_{Q}$ in Eq. (\ref{hq}) transforms through the
substitutions of $\eta (d-d^{+})=\sigma _{1}$ and $d^{+}d=(1-\sigma _{3})/2$
( $\sigma _{1,3}$ are the corresponding Pauli matrices) into the Hamiltonian
$\mathcal{H}_{S}$ of a spin $1/2$ rotating in the magnetic field $\mathbf{h}%
=(2\Delta ,0,\epsilon _{d})^{T}$ with the frequency $\omega _{0}=\sqrt{%
4\Delta ^{2}+\epsilon _{d}^{2}}$ . Then the evolution equation follows from
\begin{equation}
\partial _{t}\rho (t)=i[\rho (t),\mathcal{H}_{S}]\ .  \label{rhos}
\end{equation}%
To incorporate in it the dissipation effect due to tunneling into the empty
collector we apply the diagrammatic perturbative expansion of the S-matrix
defined by the Hamiltonian (\ref{hq}) in the tunneling amplitudes $w_{e,c}$
in the Keldysh technique \cite{konig}. This permits us to integrate out the
collector Fermi field in the following way. At an arbitrary time $t$ each
diagram ascribes indexes $a(t_{+})$ and $b(t_{-})$ of the qubit states to
the upper and lower branches of the time-loop Keldysh contour. This
corresponds to the qubit state characterized by the $\rho _{a,b}(t)$ element
of the density matrix. The expansion in $w_{e}$ produces two-leg vertices in
each line, which change the line index into the opposite one. Their effect
on the density matrix evolution has been already included in Eq. (\ref{rhos}%
). In addition, each line with index $1$ acquires two-leg diagonal vertices
produced by the electronic correlators $\langle \psi _{c}(t_{\alpha })\psi
_{c}^{+}(t_{\alpha }^{\prime })\rangle ,\ \alpha =\pm $. They result in the
additional contributions to the density matrix variation: $\Delta \partial
_{t}\rho _{10}(t)=-\Gamma \rho _{10}(t),\ \Delta \partial _{t}\rho
_{01}(t)=-\Gamma \rho _{01}(t),\ \Delta \partial _{t}\rho _{11}(t)=-2\Gamma
\rho _{11}(t),\ \Gamma =w_{c}^{2}/2$. Next, to count the electron tunnelings
into the collector we ascribe \cite{lll} the opposite phases to the
collector tunneling amplitude $w_{c}\exp \{\pm i\chi /2\}$ along the upper
and lower Keldysh contour branch, correspondingly. These phases do not
affect the above contributions, which do not mix the amplitudes of the the
different branches. Then there are also vertical fermion lines from the
upper branch to the lower one due to the non-vanishing correlator $\langle
\psi _{c}(t_{-})\psi _{c}^{+}(t_{+}^{\prime })\rangle $, which lead to the
variation affected by the phase difference as follows $\Delta \partial
_{t}\rho _{00}(t)=2\Gamma w\rho _{11}(t),\ w=\exp \{i\chi \}$. Incorporating
these additional terms into Eq. (\ref{rhos}) we come to the Lindblad quantum
master equation
\begin{eqnarray}
\partial _{t}\rho (t,w) &=&i[\rho ,\mathcal{H}_{S}]-\Gamma |1\rangle \langle
1|\rho -\Gamma \rho |1\rangle \langle 1|  \notag \\
&&+2w\Gamma |0\rangle \langle 1|\rho |1\rangle \langle 0|.  \label{lindblad}
\end{eqnarray}%
for the qubit density matrix evolution and counting the charge transfer.
Here the vectors $|0\rangle =(1,0)^{T}$ and $|1\rangle =(0.1)^{T}$ describe
the empty and filled QD, respectively. It is exact in our model with the
Hamiltonian (\ref{hq}) that takes into account many-body interaction of the
QD with the emitter Fermi sea. In our special case $\theta =-\pi $, the
Lindbladian evolution defined by the ordinary differential equation (\ref%
{lindblad}) does not have quantum memory. The physical reason for this
behavior originates from combination of the two factors: first, the instant
tunneling of electrons into the empty collector and second the perfect
screening by the emitter of the QD charge variations, which leave no traces
in the Fermi sea after each electron jump. Evolution of the system obeys the
Born-Markov description \cite{Timm}, this type of equations is well known
from the theory of open quantum systems. The first three terms on the
right-hand side of (\ref{lindblad}) generate the deterministic or no-jump
part of the evolution that can be described with a modified von Neumann
equation after inclusion of the non-Hermitian complements into $\mathcal{H}%
_{S}$ . The last term called recycling or jump operator counts the real
electron tunneling into the collector.

Solving Eq. (\ref{lindblad}) with some initial $\rho(0)$ independent of $w$
at $t=0$, we find the generating function $P(w,t)$ by taking trace of the
density matrix: $P(w,t)=Tr[\rho(w,t)]=\sum_{n=0}^\infty P_n(t)w^n$ and $%
P(w,0)=1$.

\subsection{Stationary density matrix}

Making use of the representation $\rho_{st}=[1+\sum_{l}a_{l}\sigma _{l}]/2$,
where $\sigma _{l}], \ l=1 \div 3$ are Pauli matrices, and demanding that
the right-hand side of Eq. (\ref{lindblad}) at $w=1$ vanishes after
substitution of $\rho_{st}$ in it, we find the stationary Bloch vector $%
\mathbf{a}_\infty$ with components $a_{l}$ as follows
\begin{equation}
\mathbf{a}_\infty=\frac{[2\epsilon _{d}\Delta ,-2\Delta \Gamma ,(\epsilon
_{d}^{2}+\Gamma ^{2})]^{T}}{\left( \epsilon _{d}^{2}+\Gamma ^{2}+2\Delta
^{2}\right) }\ .  \label{ainfty}
\end{equation}%
In general, an instant tunneling current $I(t)$ into the empty collector
directly measures the diagonal matrix element of the qubit density matrix
\cite{us} through their relation
\begin{equation}
I(t)=2\Gamma \rho _{11}(t,1)=\Gamma \lbrack 1-a_{3}(t)]  \label{I-t}
\end{equation}%
It gives us the stationary tunneling current as $I_{0}=2\Gamma
\Delta^{2}/(2\Delta ^{2}+\Gamma ^{2}+\epsilon _{d}^{2})$. Since in our model
$\epsilon _{d}$ is equal to the bias voltage applied to the emitter, the
current $I_{0}(\epsilon _{d})$ specifies a symmetric threshold peak in the $%
I-V$ dependence smeared by the finite tunneling rates and exhibiting the
power decrease as $\epsilon _{d}^{-2}$ far from the threshold. At $\Gamma
\gg \Delta $ this expression coincides with the perturbative results of \cite%
{5,lar} and shows the considerable growth of the maximum current $I_0(0)=
w_e^2(D/\pi \Gamma)$ due to the Coulomb interaction.

\subsection{Connection to the empty QD evolution}

Since the right-hand side of Eq. (\ref{lindblad}) is the linear
transformation of the density matrix we can write it in terms of the
superoperator acting on the Hilbert space of matrices as
\begin{equation}
\partial_t\rho = \mathcal{L}(w) \rho = \mathcal{L}(0) \rho + 2 w \Gamma
\mathcal{L}_{j} \rho \, ,  \label{lindblat2}
\end{equation}%
where the superoperator $\mathcal{L}(w) $ linear dependence on the counting
parameter $w$ is accounted for explicitly with the jump superoperator:
\begin{equation}
\mathcal{L}_{j} \rho \equiv |0 \rangle \langle 1|\rho|1 \rangle \langle 0|
\, .  \label{Ljmp}
\end{equation}%
The evolution operator for the Lindblat equation takes the following form
\begin{equation}
e^{t \mathcal{L}(w)}= e^{t \mathcal{L}(1)}+ 2(w-1)\Gamma \int_0^t d \tau
e^{(t-\tau) \mathcal{L}(w)} \mathcal{L}_{j}e^{\tau \mathcal{L}(1)} \, .
\label{eLw }
\end{equation}%
Then the Lindbladian evolution of an arbitrary initial QD state $%
\rho(0)=\rho_0$
\begin{equation}
\rho_{\rho_0}(w,t)=\rho_{\rho_0}(1,t)+ (w-1)\int_0^t d \tau \langle
I(t-\tau)\rangle_{\rho_0}\rho_E(w,\tau)  \label{rho0}
\end{equation}%
is connected to the evolution of the empty QD state
\begin{equation}
\rho_E (t,w)= e^{t \mathcal{L}(w)} |0\rangle \langle 0| \, .  \label{rhoE}
\end{equation}%
via the average transient current $\langle I(t)\rangle_{\rho_0}$. Taking
trace of both sides of (\ref{rho0}) we find relation between their
generating functions
\begin{equation}
P_{\rho_0}(w,t)=1+ (w-1)\int_0^t d \tau \langle I(t-\tau)\rangle_{\rho_0}
P(w,\tau) \, .  \label{Prho0}
\end{equation}%
\begin{figure}[tbp]
\centering \includegraphics[width=7cm]{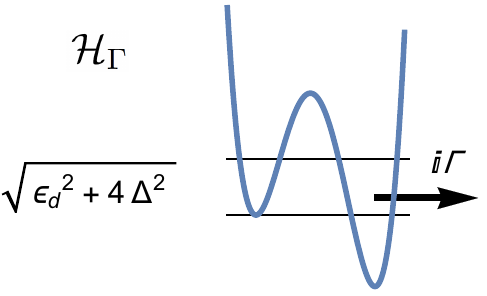}
\caption{Two level system with the coupling parameter $\Delta$ under the
bias $\protect\epsilon _{d}$. The arrow with $i \Gamma$ illustrates
tunneling from the second well.}
\end{figure}
Both relations in Eq. (\ref{rho0},\ref{Prho0}) simplify \cite{epl2} if the
initial QD state is stationary. The first one becomes
\begin{equation}
\rho_{st}(w,t)=\rho_{st}(0)+ (w-1)I_0\int_0^t d \tau \rho_E(w,\tau)
\label{rhost}
\end{equation}%
and the second after differentiating it with respect to the time can be
written as
\begin{equation}
\partial _{t}P^{st}(t,w)=(w-1)I_{0}P(t,w)\theta (t)\ ,  \label{Pst}
\end{equation}%
where the Heavyside step function $\theta (t)$ starts counting the charge
transfer at $t=0$. It is straightforward to see from Eq. (\ref{Pst}) that in
the steady process $\langle I\rangle _{st}=\partial _{w}\partial
_{t}P^{st}(t,1)=I_{0}$ and similarly one can obtain higher order moments of
steady charge transfer from this relation. Therefore, it suffices below to
focus our study on the generating function $P(t,w)$ for the process starting
from the empty QD.

\section{Generating function and Lindblad equation}

It is elucidative to derive the generating function $P(w,t)$ directly from
the Lindblad equation (\ref{lindblad}). We proceed with this here by solving
Eq. (\ref{lindblad}) perturbatively in the last term proportional $w$, which
counts the number of electron real tunneling into the collector. In the
absence of this term the evolution of qubit pure states is defined by the
evolution operator $S_{0}(t)=\exp \{-i\mathcal{H}_{\Gamma }t\}$ with the
non-Hermitian Hamiltonian:
\begin{equation}
\mathcal{H}_{\Gamma }=\Delta \sigma _{1}+(\epsilon _{d}+i\Gamma )\sigma
_{3}/2-i\Gamma /2\ .  \label{Hgamma}
\end{equation}%
The non-Hermicity leads to decrease of the amplitudes of the pure state in
the process of its evolution which is due to the electron tunneling
processes. Then the probability of observing \textquotedblleft no
tunneling\textquotedblright\ equal to the zero term $P_{0}(t)$ of the
generating function expansion reads as follows
\begin{equation}
P_{0}(t)=\sum_{a=0,1}P_{0}^{(a)}(t)=\sum_{a=0,1}|\langle a|S_{0}(t)|0\rangle
|^{2}\ .  \label{P00}
\end{equation}%
Then the generating function comes up as a perturbative series in $w$:
\begin{eqnarray}
&&P(w,t)=P_{0}(t)+\sum_{n\geq 1}(2w\Gamma )^{n}  \label{Pseries} \\
&\times
&\int_{0}^{t}dt_{1}...%
\int_{t_{n-1}}^{t}dt_{n}P_{0}(t-t_{n})P_{0}^{(1)}(t_{n}-t_{n-1})..P_{0}^{(1)}(t_{1})\ .
\notag
\end{eqnarray}%
Applying the Laplace transformation to both sides of Eq. (\ref{Pseries}) one
sums up the series and finds the generation function as follows
\begin{equation}
P(w,t)=\int_{C}\frac{dze^{zt}}{2\pi i}\frac{\check{P}_{0}(z)}{1-2w\Gamma
\check{P}_{0}^{(1)}(z)}\ ,  \label{LP}
\end{equation}%
where $\check{P}_{0}^{(a)}(z)$ stands for the Laplace transformation of $%
P_{0}^{(a)}(t)$.

\subsection{Qubit pure state evolution}

The operator $S_{0}(t)$ specifying the qubit pure state evolution between
the electron tunnelings into the collector and defined by $\mathcal{H}%
_{\Gamma }$ in Eq, (\ref{Hgamma}) takes the following explicit form
\begin{equation}
S_{0}(t)=\frac{1}{2}\sum_{\pm }e^{-(\Gamma \pm \mu \mp i\omega _{e})t/2}%
\left[ 1\mp \frac{2\Delta \sigma _{1}+(\epsilon _{D}+i\Gamma )\sigma _{3}}{%
\omega _{e}+i\mu }\right] \ ,  \label{S0}
\end{equation}%
where the parameters $\omega _{e}$ and $\mu $ are the real and imaginary
parts of $\sqrt{4\Delta ^{2}+(\epsilon _{D}+i\Gamma )^{2}}$ equal to
\begin{eqnarray}
\omega _{e} &=&\sqrt{\sqrt{\Omega ^{4}+4\Gamma ^{2}\epsilon _{D}^{2}}+\Omega
^{2}}/\sqrt{2}  \label{omegae} \\
\mu &=&\mbox{sgn}\epsilon _{D}\sqrt{\sqrt{\Omega ^{4}+4\Gamma ^{2}\epsilon
_{D}^{2}}-\Omega ^{2}}/\sqrt{2}  \label{mu} \\
\mbox{where}\ \ \Omega ^{2} &=&4\Delta ^{2}+\epsilon _{D}^{2}-\Gamma ^{2}\ .
\label{Omega}
\end{eqnarray}%
Note, that
\begin{equation}
\mu ^{2}+\omega _{e}^{2}=\sqrt{\left( \epsilon _{d}^{2}+4\Delta ^{2}-\Gamma
^{2}\right) {}^{2}+4\Gamma ^{2}\epsilon _{d}^{2}}
\end{equation}%
The two qubit states corresponding to the energies $\pm \omega _{e}/2$
possess in general different decay rates $(\Gamma \mp \mu )/2$,
respectively. The square root in Eqs. (\ref{omegae}, \ref{mu}) has its cut
along the negative real axis. Hence the oscillation frequency $\omega _{e}$
stays always positive away from the resonance and $\mbox{sgn}\mu =\mbox{sgn}%
\epsilon _{d}$ defines the relative stability of the two modes in accordance
with their relative contribution by the QD level. Say, if $\epsilon _{d}>0$
and the level contribution to the negative energy mode is bigger, the latter
decays quicker than the positive energy mode whose state locates mostly in
the emitter.

The probability $P_{0}^{(1)}(t)=|\langle 1|S_{0}(t)|0\rangle |^{2}$ of
finding the dot filled without tunneling of electrons during time $t$
follows from Eq. (\ref{S0}) as
\begin{equation}
P_{0}^{(1)}(t)=\frac{2\Delta ^{2}e^{-\Gamma t}}{\omega _{e}^{2}+\mu ^{2}}%
\left( \cosh \mu t-\cos \omega _{e}t\right) \,.  \label{P10}
\end{equation}%
In general, it is the combination of the four decaying modes of the rates $%
\Gamma \pm \mu $ and $\Gamma \pm i\omega _{e}$ due to interference in the
qubit states evolution, except for at the resonance, where either $\mu =0$
or $\omega _{e}=0$ and the number of the modes reduces to three (see below).
Its Laplace transformation is
\begin{equation}
\check{P}_{0}^{(1)}(z)=\frac{2\Delta ^{2}x}{(x^{2}+\omega
_{e}^{2})(x^{2}-\mu ^{2})}\ ,  \label{LP10}
\end{equation}%
where $x=z+\Gamma $.

Similarly the total probability $P_{0}(t)=\sum_{a}|\langle
a|S_{0}(t)|0\rangle |^{2}$ of finding no tunneling of electrons during time $%
t$ is
\begin{eqnarray}
P_{0}(t) &=&\frac{e^{-\Gamma t}}{2(\omega _{e}^{2}+\mu ^{2})}[(\omega
_{e}^{2}+\Gamma ^{2})\sum_{\pm }\left( 1\pm \frac{\mu }{\Gamma }\right)
e^{\pm \mu t}  \notag \\
&&+(\mu ^{2}-\Gamma ^{2})\sum_{\pm }\left( 1\pm \frac{i\omega _{e}}{\Gamma }%
\right) e^{\pm i\omega _{e}t}]  \label{P0}
\end{eqnarray}%
and its Laplace transformation is
\begin{equation}
\check{P}_{0}(z)=\frac{g_{E}(x)}{(x^{2}+\omega _{e}^{2})(x^{2}-\mu ^{2})}.
\label{LP0}
\end{equation}%
where $g_{E}(x)$ stands for:
\begin{equation}
g_{E}(x)=x^{3}+\Gamma x^{2}+(4\Delta ^{2}+\epsilon _{d}^{2})x+\Gamma
\epsilon _{d}^{2}\ .  \label{gE}
\end{equation}%
Note the oscillation frequency $\omega _{e}$ in Eq. (\ref{omegae}) is always
real and positive if $\epsilon _{d}\neq 0$. Hence the probabilities $%
P_{0}(t) $ and $P_{0}^{(1)}(t)$ are oscillating in time outside of the
resonance.

Meanwhile at the resonance $\epsilon _{d}=0$ the evolution operator $%
S_{0}=\exp \{-i\mathcal{H}_{\Gamma }t\}$ takes a more simple form and the
transition amplitude becomes equal to
\begin{equation}
\langle 1|S_{0}(t)|0\rangle =-i\frac{2\Delta }{\Omega }e^{-\Gamma t/2}\sin
(\Omega t/2)\ ,  \label{1S00}
\end{equation}%
where $\Omega $ is a real positive $\omega _{e}=\sqrt{4\Delta ^{2}-\Gamma
^{2}}$, if $2\Delta >\Gamma $, and it is pure imaginary $\Omega =i\mu $,
otherwise. Therefore the transition amplitude and the probability $P_{0}(t)$
are oscillating everywhere except for on the line $\Delta \in \lbrack
0,\Gamma /2]$ at $\epsilon _{d}=0$.

At the degeneracy point $2 \Delta=\Gamma$, when $\mu=\omega_e=0$, the
transition probabilities take the following forms
\begin{equation}
P^{(1)}_{0}(t)=\!\frac{\Gamma^2}{4} t^2 e^{-\Gamma t} \! , \,
P_{0}(t)=\!\left(1 + \Gamma t+ \frac{\Gamma^2}{2} t^2\right) e^{-\Gamma t}
\label{Pdegeneracy}
\end{equation}
and eventually result in an integer charge transfer statistics emitating the
fractional charge Poisson as we show below.

In the special limit $\Gamma^2+\epsilon^2_d \gg 4 \Delta^2 $ corresponding
to the perturbative calculations in \cite{5,lar} one finds that the
probabilities time decay of one mode in Eqs. (\ref{P10},\ref{P0}) becomes
much slower than the others since:
\begin{equation}
\frac{\mu^2}{\Gamma^2}=1-\frac{2 I_0}{\Gamma} +O(\frac{I^2_0}{\Gamma^2}) \,
, \ \frac{\omega_e^2}{\epsilon_d^2}=1+\frac{2 I_0}{\Gamma} +O(\frac{I^2_0}{%
\Gamma^2}) \, ,  \label{musmall}
\end{equation}
where we use $2 \Gamma\Delta^2 /(\Gamma^2+\epsilon^2_d) =I_0 \ll 2 \Gamma$,
and the probabilities converge to their single slowest mode contributions:
\begin{equation}
P^{(1)}_{0}(t)=\frac{\Delta^2 e^{-2 \Gamma\Delta^2 t/(\Gamma^2+\epsilon^2_d)}%
}{\Gamma^2+\epsilon^2_d} \, , \ P_{0}(t)=e^{-2 \Gamma\Delta^2
t/(\Gamma^2+\epsilon^2_d)} \,  \label{PtPsn}
\end{equation}
at the long enough time $t \Gamma \gg 1$. The generating function in Eq. (%
\ref{LP}) for this probability approximations reduces to the pure Poissonian
$P(w,t)=\exp\{(w-1)I_0 t\}$.

In the opposite limit $\Gamma^2+\epsilon^2_d \ll 4 \Delta^2 $ the
expressions in Eqs. (\ref{omegae},\ref{mu}) are approximated as
\begin{equation}
\omega_e^2=4 \Delta^2+\epsilon_d^2-\Gamma^2 \, , \ \mu^2=\frac{%
\epsilon_d^2\Gamma^2}{4 \Delta^2} \ll \Gamma^2 \, .  \label{deltalarge}
\end{equation}
This probability modes behavior demonstrate that in spite of the large
energy split both qubit states have the very close decay rates and both are
characterized by the approximately equal 1/2 probabilities of the QD
occupation.

\subsection{Generating function}

Substitution of the Laplace transformations $\check{P}_{0}^{1}(z)$ and $%
\check{P}_{0}^{1}(z)$ from Eqs. (\ref{LP0},\ref{LP10}) into (\ref{LP})
brings us the generation function as follows
\begin{equation}
P(t,w)=\int_{C}\frac{dze^{zt}}{2\pi i}\frac{\check{P}_{0}(z)(x^{2}+\omega
_{e}^{2})(x^{2}-\mu ^{2})}{(x^{2}+\omega _{e}^{2})(x^{2}-\mu ^{2})-4\Gamma
\Delta ^{2}wx}\ ,  \label{ptw2}
\end{equation}%
where the denominator under the integral can be re-written as
\begin{equation}
x^{4}+(4\Delta ^{2}+\epsilon _{d}^{2}-\Gamma ^{2})x^{2}-4\Delta ^{2}\Gamma
wx-\Gamma ^{2}\epsilon _{d}^{2} \equiv p_{4}(x,w)  \label{p4}
\end{equation}%
and the nominator is equal to $g_{E}(z+\Gamma )$ from Eq. (\ref{gE}).

First, we use this expression to calculate the non-zero $n$ coefficients of
the expansion of the generating function $P(w,t)=\sum P_{n}(t)w^{n}$, which
specify the time dependence of the probabilities of finding exactly $n$
electrons tunneled in the collector during time $t$.
\begin{equation}
P_{n}(t)=e^{-\Gamma t}\int_{C}\frac{dxe^{xt}}{2\pi i}Q_{n}(x)\ ,
\label{PnInt}
\end{equation}%
where $Q_{n}(x)$ are%
\begin{equation}
Q_{n}(x)=\frac{(4\Delta ^{2}\Gamma x)^{n}\left( \epsilon _{d}^{2}(\Gamma
+x)+x\left( 4\Delta ^{2}+x^{2}+\Gamma x\right) \right) }{\left( \epsilon
_{d}^{2}\left( x^{2}-\Gamma ^{2}\right) +x^{2}\left( -\Gamma ^{2}+4\Delta
^{2}+x^{2}\right) \right) ^{n+1}}.  \label{Qnl}
\end{equation}%
Note, that $Q_{0}(x)\equiv \check{P}_{0}(z)$ at $z=x+\Gamma $. Closing the
contour $C$ of the integral in Eq. (\ref{PnInt}) in the left half-plane and
counting the residues of its four degenerate poles we find for arbitrary $n$
\begin{equation}
P_{n}(t)=e^{-\Gamma t}\sum_{l=-1}^{2}res[e^{tx_{l}}Q_{n}(x_{l})];\ \ \
x_{l}=\pm \mu ,\pm i\omega _{e}  \label{Pnw0}
\end{equation}%
where $res[e^{tx_{l}}Q_{n}(x_{l})]$ are residues of the $e^{tx}Q_{n}(x)$ at $%
x=x_{l}.$ Behavior of the first five $P_{n}(t)$ is depicted in Fig. \ref%
{fig:pn}. It shows their visible $\omega _{e}$ frequency oscillations and
the exponential decay rate. Therefore, observation of the fixed number
electron tunneling permits us to extract a direct information of the qubit
evolution ruled by $\mathcal{H}_{\Gamma }$ including the energy split $%
\omega _{e}$ of the qubit states and their decay rates $\Gamma \pm \mu $.

\begin{figure}[tbp]
\centering \includegraphics[width=8cm]{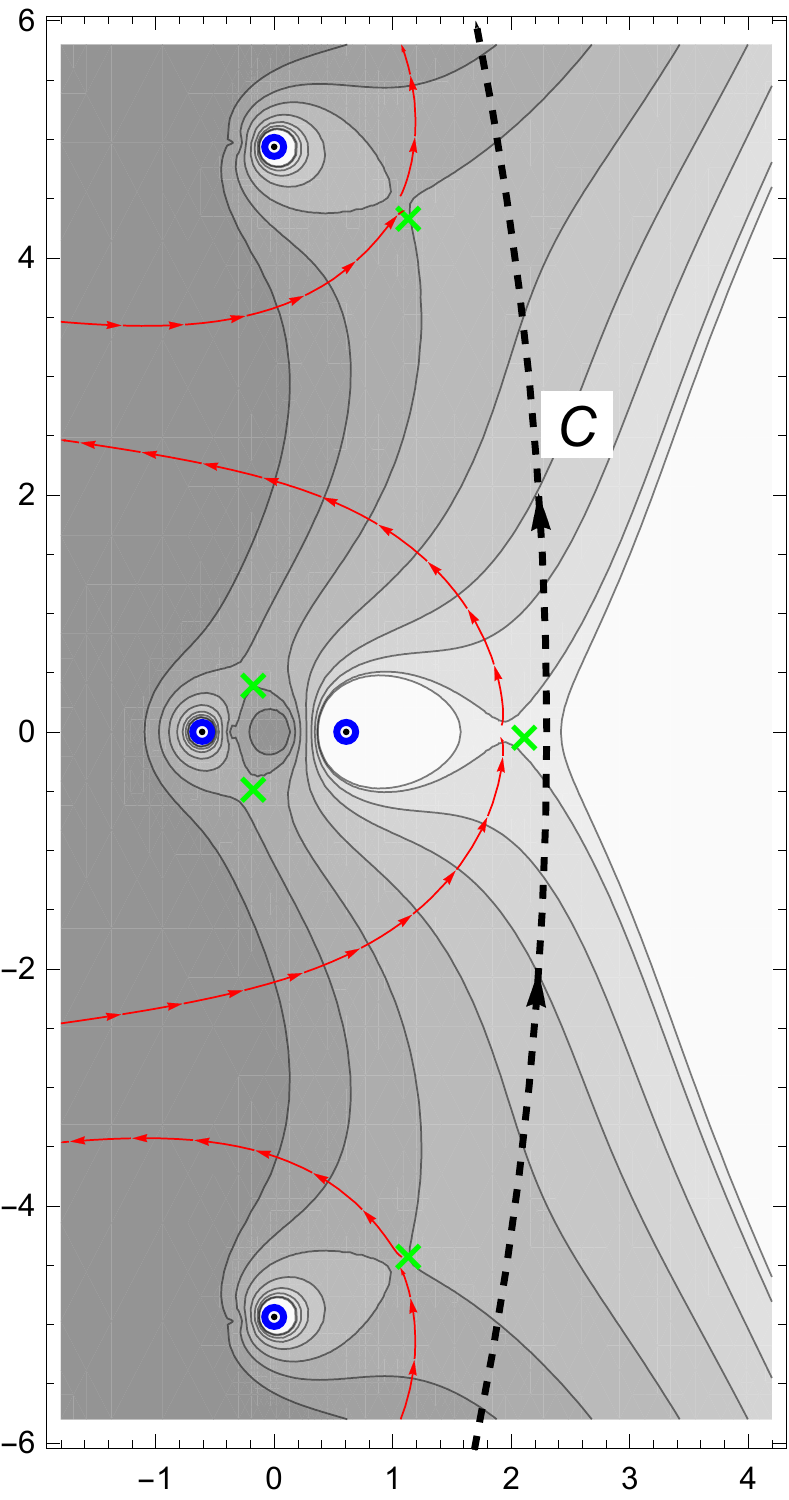}
\caption{Contour plot of Re$[S(x)]$ as a function of complex variable $x$
for $n=5$, $t=4.5,\Gamma =1$, $\protect\epsilon _{d}3$ and $\Delta =2$. The
green crosses show roots of Eq. \protect\ref{saddle_point} and the points
in the blue circles are poles of the function $Q_{n}(x)$. The thick dashed
curve is the integration countour $C$. The red curves with arrows show the
steepest descent path for the contour transformation.}
\label{fig:Cc}
\end{figure}

\begin{figure}[tbp]
\centering \includegraphics[width=8cm]{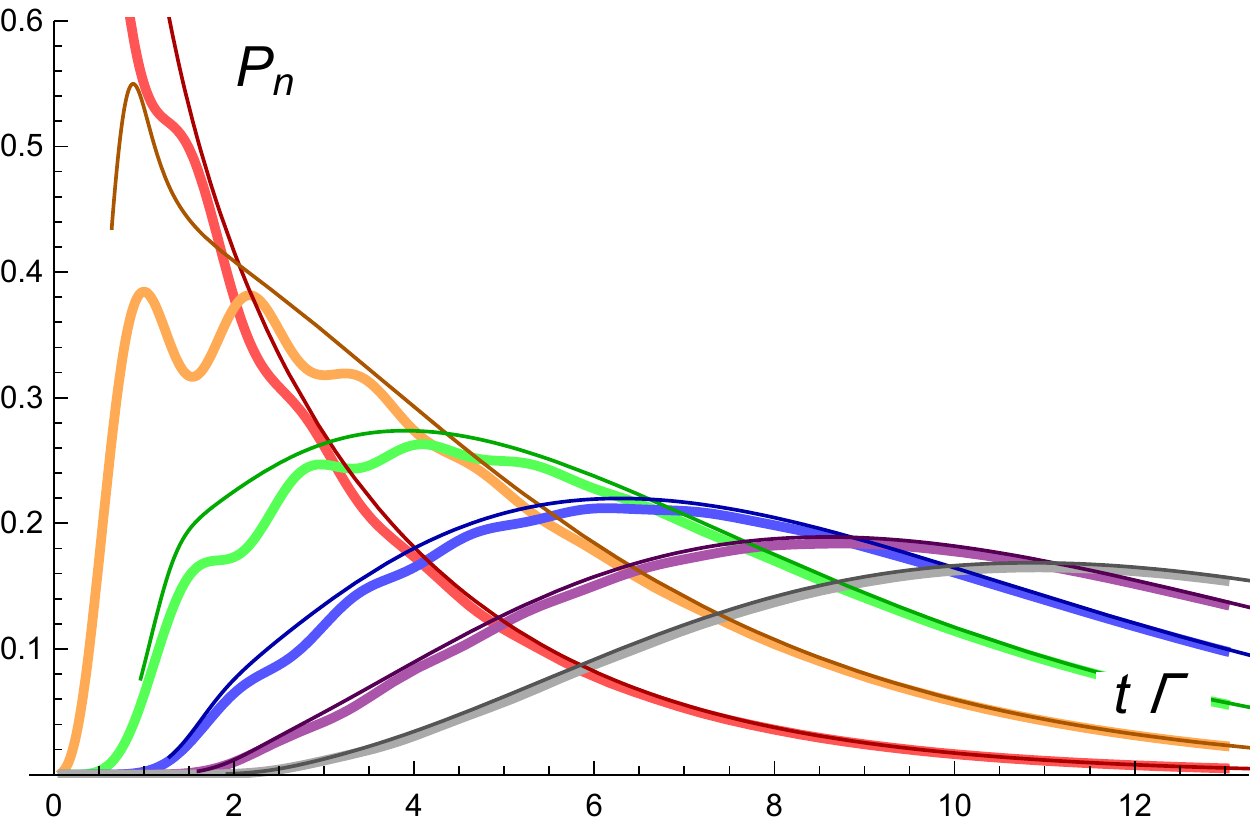}
\caption{Plot of the probabilities $P_{n}$ in Eq. \protect\ref{Pnw0} as a
function of $t$ for $\protect\epsilon _{d}/\Gamma =3$ and $\Delta /\Gamma =2$%
. The red, brown, green, blue, purple and grey lines correspond to the
parameter $n=$ 0, 1, 2, 3, 4, and 5. The thin curves of the same colors
illustrate Eq. \protect\ref{pn_saddle} for the same parameters $n$, $\protect%
\epsilon _{d}/\Gamma $ and $\Delta /\Gamma $. }
\label{fig:pn}
\end{figure}

In order to evaluate $P_n(t)$ in Eq. (\ref{PnInt}) at
large $t$ or large $n$ one can use the saddle point
approximation \cite{Fedoryuk}.%
\begin{equation}
P_{n}^{(s)}(t)=e^{-\Gamma t}\frac{1}{\sqrt{2\pi S^{\prime \prime }(x_{s})}}%
e^{x_{s}t}Q_{n}(x_{s})\ ,  \label{pn_saddle}
\end{equation}%
where $S(x)=tx+\ln [Q_{n}(x)]$ and the saddle points $x_{s}$ are defined by
the condition $S^{\prime }(x_{s})=0$. It reads as
\begin{eqnarray}
x_{s}t &=&\frac{\Gamma \epsilon _{d}^{2}-x_{s}^{2}\left( \Gamma
+2x_{s}\right) }{\left( \Gamma +x_{s}\right) \left( \epsilon
_{d}^{2}+x_{s}^{2}\right) +4\Delta ^{2}x_{s}}+  \notag \\
&&n+1\!+\frac{(n+1)2\left( \Gamma ^{2}\epsilon _{d}^{2}+x_{s}^{4}\right) }{%
\left( x_{s}^{2}-\Gamma ^{2}\right) \left( \epsilon
_{d}^{2}+x_{s}^{2}\right) +4\Delta ^{2}x_{s}^{2}}
\end{eqnarray}%
and at large $n$ yields the equation%
\begin{equation}
\frac{1}{x_{s}}+\frac{2\left( \Gamma ^{2}\epsilon _{d}^{2}+x_{s}^{4}\right)
}{x_{s}[\left( x_{s}^{2}-\Gamma ^{2}\right) \left( \epsilon
_{d}^{2}+x_{s}^{2}\right) +4\Delta ^{2}x_{s}^{2}]}=\frac{t}{n+1}\,.
\label{saddle_point}
\end{equation}%
The largest real root of this equation $x_{s0}>\mu $ corresponds to the
major saddle point, which is the left green cross shown in Fig. (\ref{fig:Cc}). It
is convenient to use $x_{s0}$ as parameter and draw parametric plot with $t$
defined by Eq. (\ref{saddle_point}) and $P_{n}(t)$ by Eq. (\ref{pn_saddle}).
The results are shown in Fig. (\ref{fig:pn}) as the thin curves of the same
color for each $n$. This approximation works well at large $t$ for any $n$
and at large $n$ for arbitrary $t,$ however it does not show the probability
oscillations. The contribution to the integral (\ref{PnInt}) that generates
oscillations comes from the two complex conjugate roots $x_{s1,2}$ of Eq. (%
\ref{saddle_point}) with positive real part. In Fig. (\ref{fig:Cc}) they are
shown as the green crosses near the poles $\pm i\omega _{e}$. Bending the integration
contour along the steepest descent paths we get two additional contributions
similar to (\ref{pn_saddle}) to the integral (\ref{PnInt}) from these saddle
points. The amplitude of the oscillations is
\begin{equation}
A_{n}^{(s)}(t)=e^{-\Gamma t}\frac{1}{\sqrt{2\pi |S^{\prime \prime }(x_{s1})|}%
}\left\vert e^{x_{s1}t}Q_{n}(x_{s1})\right\vert \ .
\end{equation}%
Near its maximum the expression (\ref{pn_saddle}) allows further
simplification and reduces to the Generalized inverse Gaussian distribution
of variable $n$ at a fixed moment of time \cite{GIG}
\begin{equation}
P_{n}^{(G)}(t)=\frac{1}{I_{0}}\frac{1}{\sqrt{2\pi n\sigma }}\ \exp [-\frac{%
(n-tI_{0})^{2}}{2n\sigma I_{0}^{2}}],  \label{GIG}
\end{equation}%
with the mean value $n_{0}=tI_{0}$ and the variance $var=8\sigma
^{2}+2t\sigma /I_{0},$ where%
\begin{equation}
\sigma =\frac{\Gamma ^{4}-2\Gamma ^{2}\Delta ^{2}+2\left( \Gamma
^{2}+3\Delta ^{2}\right) \epsilon _{d}^{2}+\epsilon _{d}^{4}+4\Delta ^{4}}{%
4\Gamma ^{2}\Delta ^{4}}  \label{sigma}
\end{equation}%
The asymptotic expression (\ref{GIG}) is a Gaussian function of time that
satisfies, in fact, the general relation
\begin{equation}
\int_{0}^{\infty }P_{n}(t)dt=Q_{n}(\Gamma )=I_{0}^{-1}\ ,  \label{int pn}
\end{equation}%
which follows from Eq. ( \ref{ptw2}) as a direct consequence of $p_{4}(x,w)$
being linear in $w$ and $p_{4}(\Gamma ,1)=0$. The integral in (\ref{int pn})
gives us a visibility time frame for observation of the fixed number
tunnelings.

\section{Current cumulants and transient extra charge}

Next, we avail of the generating function ( \ref{ptw2}) in the standard way
to obtain the average moments of the charge distribution and its cumulants.
The latter growing linearly with time are particular convenient to
characterize the long time behavior of the charge distribution, while the
formers describe the transient behavior of the charge distribution and, in
particular, the oscillatory transient current \cite{epl}.

The suitable $P(w,t)$ expression follows from calculation of the integral in
Eq. (\ref{ptw2}) by closing the contour $C$ in the left half-plane and
counting the residues of the four integrand poles defined by the roots $%
x_{l},\ l=-1\div 2$ of $p_{4}(x)$ in Eq. (\ref{p4}). which results in
\begin{equation}
P(t,w)=\sum\limits_{l=-1}^{2}q_{l}(w)\exp [(x_{l}(w)-\Gamma )t]\,.
\label{ptw1}
\end{equation}%
Here the coefficients
\begin{equation}
q_{l}(w)=\frac{\Gamma \epsilon _{d}^{2}+\epsilon _{d}^{2}x_{l}+\Gamma
x_{l}^{2}+4\Delta ^{2}x_{l}+x_{l}^{3}}{2x_{l}\left( \epsilon _{d}^{2}-\Gamma
^{2}+4\Delta ^{2}\right) +4x_{l}^{3}-4\Gamma \Delta ^{2}w}  \label{qlresi}
\end{equation}%
do not depend on time and should meet the following conditions:
\begin{equation}
\partial _{w}^{n}\sum\limits_{l=-1}^{2}q_{l}(w)=0=\partial
_{w}^{n}\sum\limits_{l=-1}^{2}q_{l}(w)x_{l}(w)|_{w=1}\,,\ n\geq 1\,.
\label{qlsum}
\end{equation}%
The first of these restrictions stems from the normalization $P(w,0)=1$,
while the second equation reflects that the process starts from the empty
state of QD, since the moments of the transferred charge are given by $%
\left\langle N^{n}(t)\right\rangle =(w\partial _{w})^{n}P(t,w)$ at $w=1$. It
also means that the sum on the right-hand side is a constant around $w=1$.

The long time behavior of the moments is determined by the term in Eq. (\ref%
{ptw1}) with the main root $x_{0}(w)$ exponent, where
\begin{equation}
x_{0}(1)=\Gamma \ \ \mbox{and}\ \ q_{0}(1)=1\,.  \label{w1}
\end{equation}%
The other exponents in Eq. (\ref{ptw1}) contribute to the transient behavior
of the transferred charge moments and specify, in particular, the transient
current time dependence $\left\langle I(t)\right\rangle $. From calculation
of the transient current in \cite{epl} we conclude that $q_{l}(1)=0$, if $%
l\neq 0$, and
\begin{equation}
\left\langle I(t)\right\rangle =x_{0}^{\prime }(1)+\sum_{l\neq
0}q_{l}^{\prime }(1)(x_{l}(1)-\Gamma )\exp [(x_{l}(1)-\Gamma )t]  \label{It}
\end{equation}%
Although the prefactor $q_{0}(w)$ at the main exponent in Eq. (\ref{ptw1})
does not contribute to the transient current it contains information of the
total charge accumulation. Indeed, integrating the right-hand side of Eq. (%
\ref{It}) over time and using the relation (\ref{qlsum}) one finds the
transient extra in the long time asymptotics of the average charge as
follows
\begin{equation}
\delta \left\langle N(t)\right\rangle =\left\langle N(t)\right\rangle
-tI_{0}\asymp q_{0}^{\prime }(1)\,,t\rightarrow \infty  \label{DeltaN}
\end{equation}%
Direct differentiation of Eq. (\ref{qlresi}) gives us the explicit
expression for the average extra charge:
\begin{equation}
q_{0}^{\prime }(1)=\frac{\Delta ^{2}(\epsilon _{d}^{2}-3\Gamma ^{2})}{%
(\epsilon _{d}^{2}+\Gamma ^{2}+2\Delta ^{2})^{2}}\ ,  \label{q0prime}
\end{equation}%
which is negative near the resonance and becomes positive if $\epsilon
_{d}^{2}$ exceeds $3\Gamma ^{2}$. As the integral of the function $%
\left\langle I(t)\right\rangle -I_{0}$ the average extra charge can be
positive only if the transient current $\left\langle I(t)\right\rangle $
varies from $\left\langle I(0)\right\rangle =0$ to $\left\langle I(\infty
)\right\rangle =I_{0}$ non-monotonically and grows bigger than $I_{0}$ at
some times. This occurs in our system because of the oscillating behavior of
the transient current as will be illustrated later on examples in Special
regimes. Similarly can be found higher cumulants of the extra charge
fluctuations, which we discuss below.

\subsection{Zero frequency current cumulants}

The leading asymptotics of $\ln P(w,t)$ at large $t$ and $w\approx 1$ is
specified by the largest root of $p_{4}$ as
\begin{equation}
\ln P(t,w)\asymp t[x_{0}(w)-\Gamma ]+\ln q_{0}(w).  \label{lnPTW}
\end{equation}%
Then $x_{0}(w)$ serves as the CGF and the reduced zero-frequency current
correlator or cumulant of the $n$th order is $\left\langle \left\langle
I^{n}\right\rangle \right\rangle =(w\partial _{w})^{n}x_{0}(w)$ at $w=1$.

\begin{figure}[tbp]
\centering \includegraphics[width=8cm]{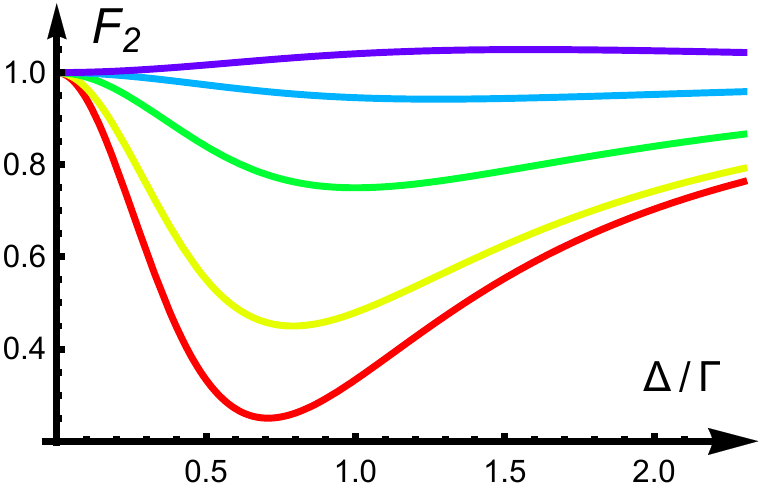}
\caption{Plot of the Fano factor $F_2$ in Eq. \protect\ref{f2} as a function
of $\Delta/\Gamma$. The red, yellow, green, light blue and blue lines
correspond to the parameter $\protect\epsilon_{d}/\Gamma=$ 0, 0.5, 1, 1.5,
and 2. }
\label{fig:I01}
\end{figure}

Since the explicit analytic expression for the root is too cumbersome, we
will calculate the cumulants $\langle \langle I^{n}\rangle \rangle $ through
the root Taylor expansion around $w=1$ in the following form:
\begin{equation}
x_{0}(\,e^{i\chi })=\Gamma +\sum\limits_{n=1}\left\langle \left\langle
I^{n}\right\rangle \right\rangle (i\chi )^{n}/n!\,.  \label{x0chi}
\end{equation}%
Normalizing the results $F_{n}=\left\langle \left\langle I^{n}\right\rangle
\right\rangle /I_{0}$ we find the Fano factor equal to:
\begin{equation}
F_{2}=1+\frac{2\Delta ^{2}(\epsilon _{d}^{2}-3\Gamma ^{2})}{(\epsilon
_{d}^{2}+\Gamma ^{2}+2\Delta ^{2})^{2}}\   \label{f2}
\end{equation}%
and its behavior is shown in Fig. (\ref{fig:I01}).

This expression shows the clear border $\epsilon _{d}^{2}=3\Gamma ^{2}$
between the sub-Poissonian current fluctuations near the resonance at $%
\epsilon _{d}=0$ and the super-Poissonian ones far from it. From comparison
of Eqs. (\ref{q0prime}) and (\ref{f2}) we conclude that
\begin{equation}
F_{2}=1+2 \delta \left\langle N(\infty )\right\rangle .  \label{f2N}
\end{equation}%
The reason for this seemingly accidental relation between the Fano factor
and the average extra charge will be clarified below. The Fano factor $F_{2}$
reaches its minimum $F_{2}=0.25$ at $\epsilon _{d}=0$ and $\Delta =\Gamma /%
\sqrt{2}$ and it asymptotically approaches its maximum $F_{2}\rightarrow
1.25 $ as $\epsilon _{d}=\sqrt{2}\Delta \rightarrow \infty $.

\begin{figure}[tbp]
\centering \includegraphics[width=8cm]{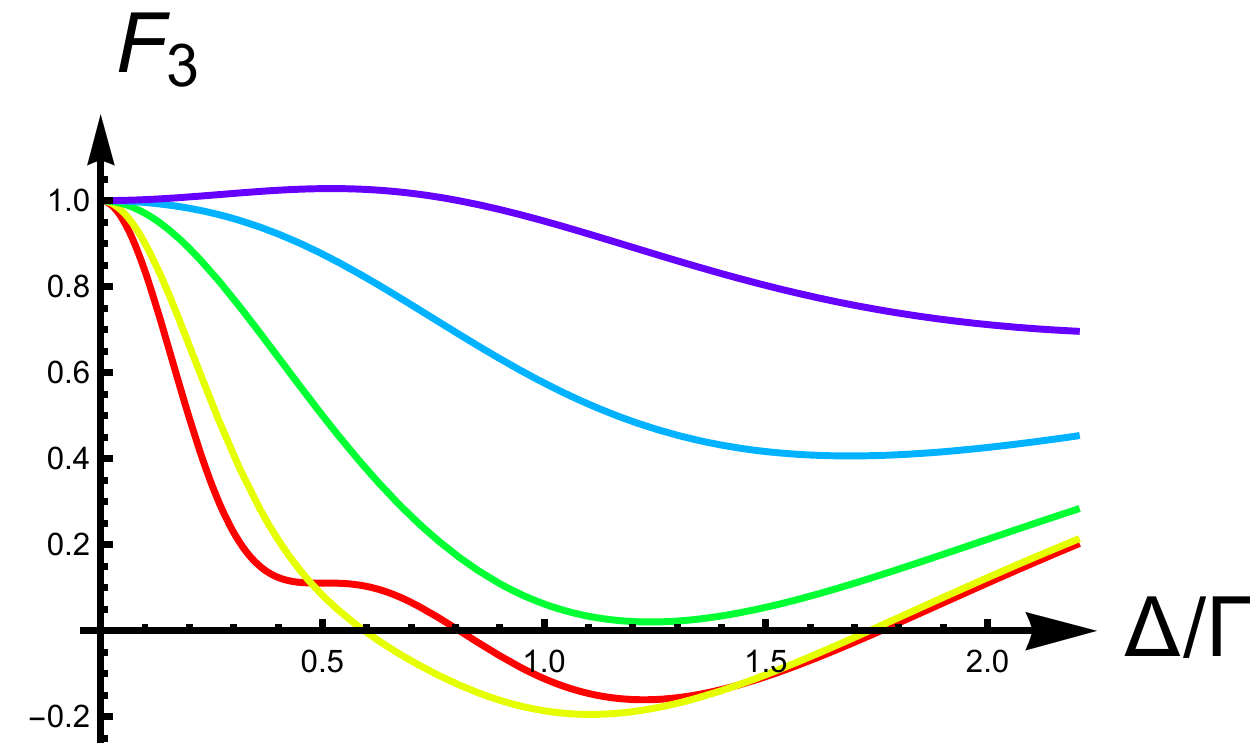}
\caption{Plot of the skewness $F_3$ in Eq. \protect\ref{f3} as a function of
$\Delta/\Gamma$. The red, yellow, green, light blue and blue lines
correspond to the parameter $\protect\epsilon_{d}/\Gamma=$ 0, 0.5, 1, 1.5,
and 2. }
\label{fig:I02}
\end{figure}

\begin{figure}[tbp]
\centering \includegraphics[width=7cm]{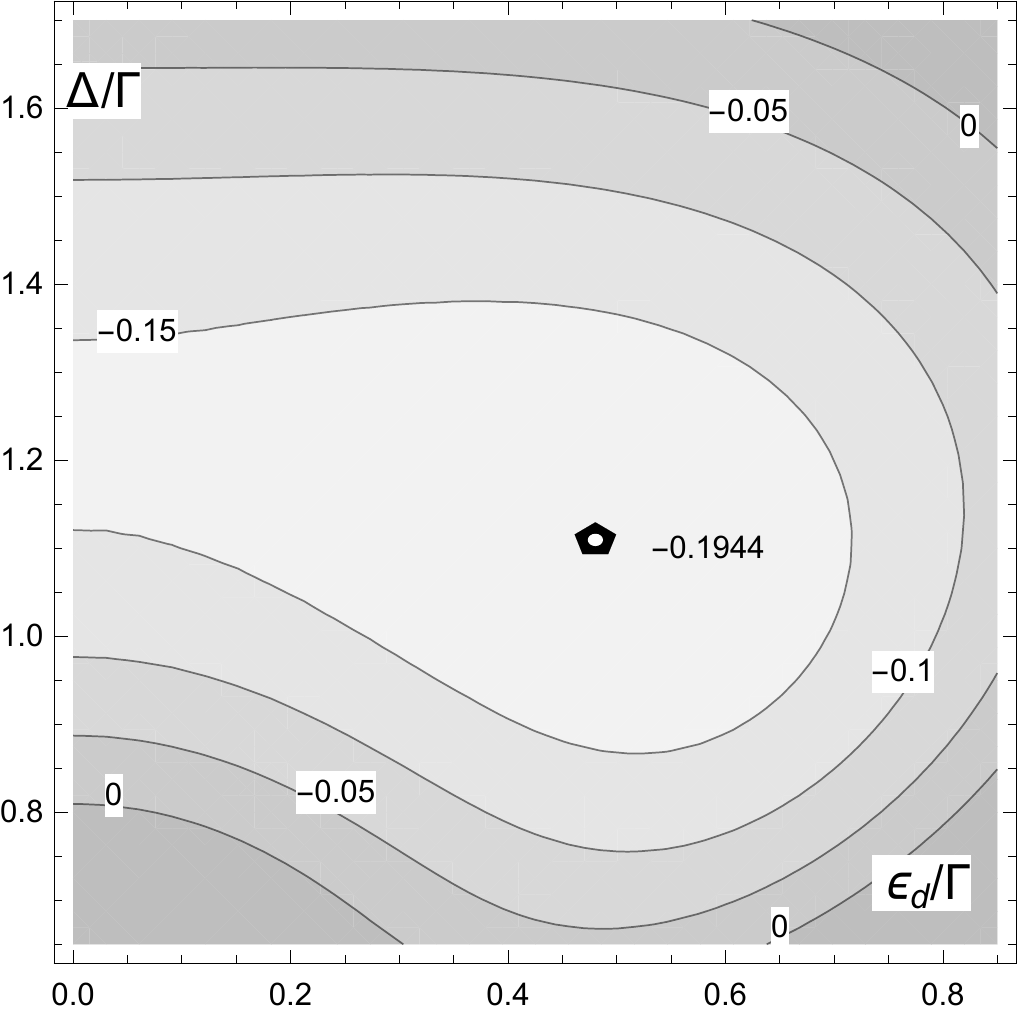}
\caption{Plot of the area where the skewness $F_3$ in Eq. \protect\ref{f3}
is negative as a function of $\Delta/ \Gamma$ and $\protect\epsilon_{d}/
\Gamma$. The black pentagon with white point indicates its absolute minimum.
}
\label{fig:I03}
\end{figure}

The third order normalized commulant called skewness is equal to
\begin{eqnarray}
\!\!\!\!\!\!\!&&\!\!\!\!\!F_3=1+\frac{48 \left(4 \Gamma ^4 \Delta ^4+4
\Gamma ^2 \Delta ^6+\Delta ^8\right)}{\left(\Gamma ^2+\epsilon _d^2+2 \Delta
^2\right){}^4}+  \label{f3} \\
\!\!\!\!\!\!\!&&\!\!\!\!\! \frac{6 \Delta ^2 \left(\epsilon _d^4-3 \Gamma
^4-22 \Gamma ^2 \Delta ^2-2 \left(\Gamma ^2-\Delta ^2\right) \epsilon _d^2-4
\Delta ^4\right)}{\left(\Gamma ^2+\epsilon _d^2+2 \Delta ^2\right){}^3} \ .
\notag
\end{eqnarray}
and depicted in Figs. (\ref{fig:I02},\ref{fig:I03}). Among the special
features of its behavior we observe a small parametric area, where the
skewness is negative, and also appearance of a plateau in its parameter
dependence at the degeneracy point $2 \Delta=\Gamma, \epsilon _d=0 $
characterized by the transition probabilities in Eq. (\ref{Pdegeneracy}).

\subsection{Transient extra charge fluctuations and Fano factors}

As follows from Eq. (\ref{lnPTW}) the long time asymptotics of the cumulants
of the transferred charge statistics:
\begin{equation}
\left\langle \left\langle N^{n}(t)\right\rangle \right\rangle \asymp
tI_{0}F_{n}+(w\partial )^{n}\ln q_{0}(w)|_{w=1}.  \label{nTW}
\end{equation}%
contains besides the terms growing linearly in time and defined by the Fano
factors $F_{n}$, the additional non-universal contributions due to the
transient extra charge accumulation depending on the initial state of QD.
The extra charge cumulants are defined by their CGF $\ln q_{0}(w) $ and
could be formally considered as a result of an independent additional charge
transfer process. This process, however, does not make a clear physical
sense as the extra charge second order cumulant
\begin{eqnarray}
\!\!\!\!\!\!\! &&\!\!\!\!\!\delta \left\langle \left\langle
N^{2}\!\right\rangle \right\rangle= \frac{d[w(d\ln[q_{0}])}{dw^{2}}=\frac{%
I_{0}{}^{4}\left( \epsilon _{d}^{2}-3\Gamma ^{2}\right) \left( \Gamma
^{2}+\epsilon _{d}^{2}\right) {}^{2}}{16\Gamma ^{4}\Delta ^{6}}  \notag \\
\!\!\!\!\!\!\! &&\!\!\!\!\!+\frac{I_{0}{}^{4}\left( 25\Gamma ^{4}-54\Gamma
^{2}\epsilon _{d}^{2}+\epsilon _{d}^{4}\right) }{16\Gamma ^{4}\Delta ^{4}}-%
\frac{I_{0}{}^{4}\left( 7\Gamma ^{2}+3\epsilon _{d}^{2}\right) }{4\Gamma
^{4}\Delta ^{2}}  \label{deltaN2}
\end{eqnarray}
is not always non-negative.

The expression for the long time asymptotics of the transferred charge
cumulants analogous to Eq. ( \ref{nTW}) can also be written for the steady
evolution process starting in the stationary QD state. However, in this case
there is no additional average charge accumulation $\left\langle
N(t)\right\rangle _{st}=I_{0}t$ and hence $q_{st,0}^{\prime }(1)=0$. Making
use of the general relation between the generating functions for both
processes we substitute their asymptotics (\ref{lnPTW}) into Eq. ( \ref{Pst}%
) and find that
\begin{equation}
x_{0}(w)-\Gamma =(w-1)I_{0}q_{0}(w)/q_{st,0}(w).  \label{x0q}
\end{equation}%
Taking the second derivative of this equation with respect to $i\chi $ at $%
w=\exp (i\chi )=1$ we come to the relation $F_{2}=1+2q_{0}^{\prime }(1)$
derived earlier in (\ref{f2N}) between the Fano factor and the extra
transient average charge $\delta \left\langle N(\infty )\right\rangle $ as
defined in Eq. (\ref{DeltaN}). This relation explains, in particular, that
the super Poissonian current fluctuations in this system occurs due to an
excess of the average transient charge accumulated in the tunneling process
initiated in the empty state of QD, contrary to the more common sub
Poissonian current shot noise, which happens if there is a deficit of this
average charge. Moreover, since the excessive average charge needs a
non-monotonous time dependence of the transient current, the emergence of
the super Poissonian noise is a finger-print of the qubit coherent dynamics
in the system and an oscillating behavior of the transient current.

Taking the third derivative of Eq. ( \ref{x0q}) we can write the third order
Fano factor in the following form
\begin{equation}
F_{3}=1+\frac{3}{4}[F_{2}^{2}-1]+3[\delta \left\langle \left\langle \!
N^{2}\!\right\rangle \right\rangle -\!\delta\left\langle \left\langle
\!N^{2}\!\right\rangle \right\rangle _{st}]\ ,  \label{f3N2}
\end{equation}%
which relates the skewness to the difference in the transient extra charge
fluctuations in the two processes. Although the second term on the
right-hand side in (\ref{f3N2}) is negative for the sub-Poissonian noise,
the whole skewness also becomes negative only when the second order cumulant
defined by $\delta \left\langle \left\langle \! N^{2}\!\right\rangle
\right\rangle _{st}=\left\langle \left\langle \!N^{2}(t)\!\right\rangle
\right\rangle _{st}-I_{0}F_{2}t$ at large $t$ and characterizing the extra
fluctuations of the charge accumulated in the steady evolution process is
much bigger than the corresponding cumulant $\delta \left\langle
\left\langle \!N^{2}\!\right\rangle \right\rangle =\left\langle \left\langle
\!N^{2}(t)\!\right\rangle \right\rangle -I_{0}F_{2}t$ at large $t$ of the
extra charge fluctuations in the case of the evolution process starting from
the empty QD. The area where $F_{3}<0$ is shown on Fig.\ref{fig:I03}.

\subsection{Special regimes}

From the $p_{4}(x)$ expression in the integrand denominator in Eq. (\ref%
{ptw2}) we find the linear in $w$ dependence of the roots as $x_{0,2}=\pm
\mu +x_{0}^{\prime }(0)w$ and $x_{\pm 1}(w)=\pm i\omega _{e}-x_{0}^{\prime
}(0)w$, where
\begin{equation}
x_{0}^{\prime }(0)=\frac{2\Delta ^{2}\Gamma }{\omega _{e}^{2}+\mu ^{2}}\
\label{Deltax}
\end{equation}%
is well defined except for at the degeneracy point.

The linear root approximations can be extended up to $w=1$ if $x_{0}^{\prime
}(0)\ll |\mu |,\omega _{e}$. This limits their applicability to $\Gamma
^{2}+\epsilon _{d}^{2}\gg 4\Delta ^{2}$, where $x_{0}^{\prime }(0)=2\Delta
^{2}\Gamma /(\Gamma ^{2}+\epsilon _{d}^{2})\approx I_{0}$. Making use of
these root approximations in calculation of the Laplace transformation in
Eq. (\ref{ptw2}) we find
\begin{eqnarray}
\!\!\!\!\!\!\! &&\!\!\!\!\!P(t,w)=\left[ 1+\frac{\Delta ^{2}(w-1)\left(
\epsilon _{d}^{2}-3\Gamma ^{2}\right) }{\left( \Gamma ^{2}+\epsilon
_{d}^{2}\right) {}^{2}}\right] e^{tI_{0}(w-1)}  \label{ptwdelta0} \\
\!\!\!\!\!\!\! &&\!\!\!\!+\frac{2\Gamma ^{2}\Delta
^{2}(w-1)e^{-(I_{0}w+\Gamma )t}}{\left( \Gamma ^{2}+\epsilon _{d}^{2}\right)
{}^{2}}\sum_{\pm }\left[ 1\pm i\frac{\epsilon _{d}}{\Gamma }\right] e^{\pm
it\epsilon _{d}(1+I_{0}/\Gamma )}  \notag \\
\!\!\!\!\!\!\! &&\!\!\!\!-\frac{\Delta ^{2}(w-1)e^{t\left(
I_{0}(w+1)-2\Gamma \right) }}{\Gamma ^{2}+\epsilon _{d}^{2}}%
+O(I_{0}^{2}/\Gamma ^{2})\,.  \notag
\end{eqnarray}%
The main exponent in Eq.( \ref{ptwdelta0}) coincides with the Poissonian
defined by the single long living mode of the probability $P_{0}(t)$ in (\ref%
{PtPsn}). The prefacot at the main exponent shows that due to the charge
accumulated in the transient regime of $\Gamma t<1$ this Poissonian
eventually acquires an excessive charge of one binomial attempt in the long
time limit for $\epsilon _{d}^{2}>3\Gamma ^{2}$ and the lack of it,
otherwise. The first regime is the super-Poissonian in agreement with Eq. (%
\ref{f2}) and the other is the sub-Poissonian. To clarify the physical
mechanism of transition between these two regimes we exploit the generating
function asymptotics (\ref{ptwdelta0} ) to calculate the deviation of the
transient current from its long-time stationary limit
\begin{equation}
\left\langle \delta I(t)\right\rangle =I_{0}[e^{-2\Gamma t}-2\cos (\epsilon
_{d}t)e^{-\Gamma t}+O(I_{0}/\Gamma )]\ ,  \label{DeltaI}
\end{equation}%
which shows how increase of the frequency of the current oscillations in
comparison to the decay rate diminishes contribution of the oscillating term
on the right-hand side of (\ref{DeltaI} ) into the extra charge accumulation
and makes the average extra charge excessive.

In the opposite regime of large $\Delta $, where $\Gamma^2+\epsilon^2_d \ll
4 \Delta^2 $, the roots dependence on $w$h starting from their initial
values in Eq. (\ref{deltalarge}) can be found with increase of $w$ in $%
p_4(x) $ in the following way. $x_\pm(w)$ due to their large imaginary parts
undergo just the linear shift as $x_\pm(w)=\pm i \omega_e -\Gamma w/2$.
Meanwhile the $x_{0,2}$ dependences are essentially non-linear and at $w
\approx 1$ are given by
\begin{equation}
x_2=- \frac{\Gamma \epsilon_d^2}{4 \Delta^2 w} \, , \ x_0=\frac{\Gamma w}{4
\Delta^2}(4 \Delta^2-\epsilon_d^2+(1-w^2)\Gamma^2)-x_2 \, .  \label{x20}
\end{equation}
Making use of Eq. (\ref{qlresi}) with these roots approximation under
condition $\epsilon_d^2\gg \Gamma^2 $ we find the asymptotics of the
generating function in this super Poissonian regime as
\begin{eqnarray}
\!\!\!\!\!\!\!\!\! &&\!\!\!\!\!P(t,w)=-\frac{\epsilon _{d}^{2}(w-1)
e^{-\Gamma (1+\frac{\epsilon _{d}^{2}}{4\Delta ^{2}w})t} }{4w^{2}\Delta ^{2}}%
+  \label{Plargedelta} \\
\!\!\!\!\!\!\!\!\! &&\!\!\!\! \left( 1+\frac{\epsilon _{d}^{2}(w-1)}{%
4\Delta^{2}w^{2}}\right) e^{\Gamma \left(1-\frac{\epsilon_{d}^{2}}{ 4\Delta
^{2}} \right) (w-1)t + \frac{\Gamma \epsilon_{d}^{2}}{ 4\Delta ^{2}} \left(%
\frac{1}{w}-1 \right) t} -  \notag \\
\!\!\!\!\!\!\!\!\! &&\!\!\!\!\!\frac{4\Gamma \Delta ^{2}(w-1)\sin \left( t%
\sqrt{\epsilon _{d}^{2}+4\Delta ^{2}}\right) e^{-\Gamma \left( \frac{2\Delta
^{2}w}{\epsilon _{d}^{2}+4\Delta ^{2}}+1\right)t }}{\left( \epsilon
_{d}^{2}+4\Delta ^{2}\right) {}^{3/2}}  \notag
\end{eqnarray}
The long-time behavior of the generating function specified by the leading
exponent in (\ref{Plargedelta} ) presents the total tunneling process as a
combination of the two independent processes. Those are the main Poissonian
of electrons tunneling characterized by the tunneling rate $\Gamma \left[%
1-\epsilon_{d}^{2}/(4\Delta ^{2})\right]$ and its weak counterpart
Poissonian of holes tunneling with the small rate $\Gamma
\epsilon_{d}^{2}/(4\Delta ^{2})$, which develops as the level position
deviates from the resonance. This makes the total process super Poissonian
since the total average current comes as difference of the tunneling rates,
while the total noise is their sum.

The deviation of the transient current from its long-time stationary limit
follows from (\ref{Plargedelta}) as
\begin{equation}
\left\langle \delta I(t)\right\rangle =\frac{\Gamma \epsilon
_{d}^{2}e^{-\Gamma t(1+\frac{\epsilon _{d}^{2}}{4\Delta ^{2}})}}{4\Delta ^{2}%
}-\frac{4\Gamma \Delta ^{2}e^{-\Gamma t(\frac{3}{2}-\frac{\epsilon _{d}^{2}}{%
8\Delta ^{2}}})}{\omega _{e}^{2}}\cos (\omega _{e}t)\ .  \label{DeltaI2}
\end{equation}%
Its integration over time shows that high frequency oscillations of the
second term make the first term contribution into the average extra charge
prevail with the result: $\left\langle \delta N(\infty )\right\rangle
=(\epsilon _{d}^{2}-3\Gamma ^{2})/(4\Delta ^{2})$.

At the resonance we find from Eq. (\ref{p4} ) that $p_{4}(x,w)=xp_{3}(x)$,
where
\begin{equation}
p_{3}(x)=x^{3}+(4\Delta ^{2}-\Gamma ^{2})x-4\Delta ^{2}\Gamma w  \label{p3}
\end{equation}%
and besides the root $x_2=0$ the three other roots read \cite{Jacobson} as
follows
\begin{equation}
x_{l}=\Gamma \sum_{\pm } e^{\pm \frac{2 \pi il}{3}} \left( 2\Delta _{\Gamma
}^{2}w\pm \sqrt{4\Delta _{\Gamma }^{4}w^{2}+[(4\Delta _{\Gamma
}^{2}-1)/3]^{3}}\right) ^{\frac{1}{3}}  \label{root}
\end{equation}%
where $\Delta _{\Gamma }=\Delta /\Gamma $. The cumulant generating function
is
\begin{equation}
\ln P(t,w)\asymp t(x_{0}-\Gamma )+\ln \frac{4\Gamma ^{2}\Delta
_{\Gamma}^{2}+\Gamma x_{0}+x_{0}^{2}}{\Gamma ^{2}\left( 4\Delta
_{\Gamma}^{2}-1\right) +3x_{0}^{2}}  \label{q00}
\end{equation}%
In both limits $\Delta _{\Gamma }\ll 1$ and $\Delta _{\Gamma}\gg 1 $ it
takes the Poissonian form $\ln P(t,w)/t\asymp I_0 \lbrack w-1]$ with the
average current $I_{0}=2\Delta ^{2}/\Gamma $ and $I_{0}=\Gamma $,
respectively.

At the degeneracy point of the qubit modes when $2 \Delta=\Gamma, \epsilon
_d=0 $, the four roots of $p_4(x,w)$ follow from Eq. (\ref{p3}) as $%
x_{l}=e^{i2\pi l/3} \Gamma w^{1/3},x_{2}=0$. With $g_{E}(x)$ defined in Eq. (%
\ref{gE}) the generating function comes up after taking the Laplace
transformation integral in the form:
\begin{equation}
P(t,w)\!=\!\!\!\sum\limits_{l=-1}^{1}\frac{(w+e^{i\frac{2\pi }{3}l}w^{\frac{1%
}{3}}+e^{i\frac{4\pi }{3}l}w^{\frac{2}{3}})e^{\Gamma t(e^{i\frac{2\pi }{3}%
l}w^{\frac{1}{3}}-1)}}{3w}.  \label{ptwdg}
\end{equation}%
For $w$ in the sector around the real positive axis this function converges
at large time to the Poissonian of the fractional charge $1/3$ modified by
an independent tunneling of one and two fractional holes, which leads to the
$1/3$ deficit of the average Poissonian charge. Since all zero-frequency
current cumulants are defined by the generating function asymptotics at $w=1$
and large $t$ (see below) they all coincides with the Poissonian cumulants
equal $\Gamma /3^{n}$ in the $n$th order as if we observe the fractional
charge tunneling process.

However there is no fractional charge tunneling in the complete generating
function in Eq. (\ref{ptwdg}) because the $2 \pi$ periodicity with respect
to the $w$ phase guaranies that $P(t,w)$ remains the integer function of $w$%
. Calculating its coefficients as
\begin{equation}
P_n(t) \!=\left(\frac{(\Gamma t)^{3n}}{3n!} + \frac{(\Gamma t)^{3n+1}}{%
(3n+1)!} + \frac{(\Gamma t)^{3n+2}}{(3n+2)!} \right) e^{-\Gamma t}
\label{Pntwdg}
\end{equation}%
we find the explicit expansion of the complete generating function in the
following form:
\begin{equation}
P(t,w)\!=e^{-\Gamma t} \sum\limits_{m=0}^{\infty} \frac{(\Gamma t)^m}{m!}
w^{[m/3]} \, ,  \label{seriesptwdg}
\end{equation}%
where $[x]$ denotes the integer part of $x$ or its antie function. It is a
reduced Poissonian distribution due to unsuccessful tunneling attempts by
the fractional charges.

\section{Conclusion}

Tunneling of spinless electrons through an interacting resonant level of a
QD into an empty collector has been studied in the especially simple, but
realistic system, in which all sudden variations in charge of the QD are
effectively screened by a single tunneling channel of the emitter. This
system has been described \cite{epl} with an exactly solvable model of a
dissipative two-level system called qubit. Its matrix element $\Delta $ of
the coupling between the two-level states is equal to the bare emitter
tunneling rate $\Gamma _{e}$ renormalized by the large factor $\sqrt{D/(\pi
\Gamma _{e})},$ whereas the damping parameter $\Gamma $ coincides with the
tunneling rate into the collector.

The exact solution to this model was earlier used to demonstrate that the
coherent qubit dynamics expected in the FES regime should manifest
themselves in an oscillating behavior \cite{epl} of the average collector
transient current in the wide range of the model parameters and also through
the resonant features of the \textit{a.c.} response \cite{prb}, though the
experimental observation of these manifestations could be a difficult
experimental task. Therefore, in this work we have studied more relevant
electron transport characteristics to the modern experiments including Fano
factor of the second \cite{8} and third (skewness) orders \cite{shovkun,
third}. In particular, we have clarified a possible mechanism leading to
appearance of the sub-Poisson and super Poisson shot noise of the tunneling
current as it has been observed in the recent experiments \cite{noise,noise2}
in the FES regime.

In this work we have used the method of full counting statistics to
calculate the generating function of the distribution of charge transferred
in process of the empty QD evolution, which is governed by the generalized
Lindblat equation. This equation describes the whole process as a succession
of time periods of the non-Hermitian Hamiltonian qubit evolution randomly
interrupted by the electron tunneling jumps from the occupied QD into the
empty collector. The qubit density matrix evolution during each of these
periods has been described as a four mode process, two modes of which are
oscillating with opposite frequencies and the same damping rate $\Gamma$
about everywhere except for at the exact resonance $\epsilon_{d}=0$ and $%
\Gamma > 2 \Delta $ . As a result the time dependent probabilities $P_n(t)$
to have a certain number $n$ of electrons tunneled into the collector, which
are determined by the matrix element of the density matrix undergoing the
non-Hermitian Hamiltonian evolution, are also oscillating except for the
same infinitely narrow parametric area. These oscillations are better
visible at small time and therefore for $P_n(t)$ with small number $n$,
since the slowest damping mode is not oscillating. Note, however, that the
frequency of these oscillations is different from the one of the transient
current: Both are the transformations of the two-level energy split of the
isolated qubit by the dissipation, though the first one accounts for
expectation for the electron tunneling into the collector, but without its
real occurrence, meanwhile the second one is due to both effects.

The four modes of the Hamiltonian evolution of the qubit density matrix lead
to a general representation of the generating function as a sum of the four
exponents with linear in time arguments, which are multiplied by the
exponent prefactors. The long time behavior of this function with the
counting parameter $w \approx 1$ is determined by the leading exponent term.
Its logarithm gives us the long time asymptotics of the CGF consisting of
the two parts, which describe two independent contributions into the
transferred charge fluctuations. The part linearly growing in time defines
the zero-frequency current cumulants and has been used to calculate the Fano
factor and the skewness. It does not depend on the initial state of the QD
and hence on the transient evolution behavior. Contrary, the other part
given by the prefactor logarithm depends on the QD initial state and has
been used as the CGF of the transient extra charge fluctuations.

Our calculation of the Fano factor has shown emergence of the sub Poissonian
behavior of the current fluctuations near the resonance which changes into
the super Poissonian as the level energy moves out of the resonance and $%
\epsilon_{d}^{2}>3 \Gamma^{2}$. On the other hand, from our consideration of
the extra charge CGFs we have found the simple linear relation between the
Fano factor and the average transient extra charge accumulated during the
empty QD evolution. It explains that the sub and super Poissonian steady
current statistics correspond to the transient accumulation of the negative
and positive average extra charge, respectively. Moreover, the positive
average extra charge can be accumulated only if the transient current is
non-monotonous in time. Therefore, emergence of the super Poissonian steady
current fluctuations signals an oscillating behavior of the transient
current and the qubit coherent dynamics according to this model of the FES.

We have also calculated the skewness and found that it changes its sign and
becomes negative in the small area near the resonance, where $%
|\epsilon_{d}|< \Gamma$ and $0.6\lesssim \Delta/\Gamma \lesssim 1.8 $. We have
understood this behavior through comparison of the extra charge CGFs for the
QD evolutions starting from its empty and stationary states, which has
related the skewness to the difference between the two extra charge
cumulants of the second order characterizing difference between the total
charge noise in these two processes. This relation has shown that the
skewness becomes negative in the sub Poissonian regime, if the total charge
noise developed in the stationary state evolution is much bigger than the
one in the evolution of the empty state.

These relations between the steady current fluctuations and the extra charge
accumulation have been illustrated with particular examples of the
generating functions in the special regimes. The two generating functions
have been calculated asymptotically in the regimes when amplitude of the
qubit two-level coupling is much smaller than the collector tunneling rate
or the absolute value of the QD level energy and in the opposite limit when
the amplitude is much larger than both of them. Accumulation of the extra
charge in these regimes is illustrated with the corresponding transient
current behavior.

We have also calculated the generating function at the special point $\Gamma
=2\Delta $ at the resonance, when the two qubit levels energies including
their imaginary parts are equal. We find that in this special case it takes
the $1/3$ fractional Poissonian form, where all probabilities of tunneling
of the fractional charges mean tunneling of the charges integer parts. The
large time limit of this function, nontheless, coincides with the true $1/3$
fractional Poisson. This example underlines that observation of the
fractional charge in the Poissonian shot noise is necessary, but not
sufficient to prove its real tunneling.

We have performed our calculations in dimensionless units with $\hbar =1$
and $e=1$. In order to return to the SI units the current $I_{0}$ should also
include the dimensional factor $e^{2}/\hbar \approx 2.43\cdot 10^{-4}$S, if $\Gamma ,$ $%
\Delta ,$ and $\epsilon _{d}$ are measured in volts. In the experiments \cite%
{lar,lar1} the collector tunneling rate is $\Gamma \approx 0.1meV$ and the
coupling parameter $\Delta \approx 0.016meV$.  To observe the special  regime
of  Eqs.  (\ref{Pdegeneracy},\ref{seriesptwdg})
one can increase the collector barrier width to obtain a heterostructure
with $\Delta =0.016meV$ and $\Gamma=2\Delta  \approx0.032meV$.
Its stationary current at the resonance is $I_{0}=2.6\,nA$ and
the zero frequency spectral density of the current noise measured in
experiments as $S_0=2 |e| I_0 F_2$ with the above dimentional $I_0$ and
$F_2$ from Eq. (\ref{f2} )
is $S_{0}\approx 2.76\cdot 10^{-28}A^{2}/Hz$.
With increase of $|\epsilon _{d}|$ the current $%
I_{0}$ is decreasing, whereas $S_{0}$ grows up to its maximum $S_{0}\approx 4.2\cdot
10^{-28}A^{2}/Hz$ at $|\epsilon _{d}|\approx 0.027meV$.
At larger $|\epsilon_{d}|$ the current shot noise becomes super Poissonian
with its zero frequency spectral density approaching $S_{0}\approx 2.5|e|I_{0}$.
According to  Ref.  \cite{epl2} the finite frequency spectral density $S_{\omega }$
varies less then 20\% if $\omega <\Gamma /(2\hbar )$. Since the frequency
corresponding to the above  value of $\Gamma $ is $\omega_\Gamma \sim
5\cdot 10^{10}s^{-1}$, we can consider
the frequency $\omega$ low enough to evaluate $S_0$, if $\omega/(2\pi)$ is below $4\,GHz.$

\begin{acknowledgments}
The work was supported by the Leverhulme Trust Research Project Grant
RPG-2016-044 (V.P.) and Russian Federation STATE TASK No 075-00475-19-00
(I.L.).
\end{acknowledgments}

\end{document}